\documentclass[11pt,twoside]{article}
\usepackage{asp2010}

\usepackage{color}

\aspvoltitle{UP: Have Observations Revealed a Variable
Upper End of the Initial Mass Function?\\
June 20-25, 2010, Sedona, Arizona, USA 
}

\resetcounters

\begin{document}
\markboth{C. Clarke \& Th. Maschberger}{Cluster mass dependent truncation of the upper IMF}

\title{Cluster mass dependent truncation of the upper IMF: evidence from
observations and simulations}
\author{C. Clarke \& Th. Maschberger $^1$
\affil{$^1$Institute of Astronomy, Madingley Road, CAMBRIDGE, U.K., CB3 OHA}
}

\begin{abstract}
We attempt  to evaluate whether the integrated galactic IMF (IGIMF) is
expected to be steeper than the IMF within individual clusters through
direct evaluation of whether there is a {\it systematic} dependence
of maximum stellar mass on cluster mass. We show that the result
is sensitive to  observational selection biases and requires an accurate
knowledge of cluster ages, particularly in more populous clusters. At face
value there is no compelling evidence for  non-random selection of stellar
masses in low mass clusters but there is arguably some evidence
that the  maximum stellar mass is anomalously low (compared
with the expectations of random mass selection) in clusters
containing more than several thousand stars. Whether or not this effect
is then imprinted on the IGIMF then depends  on the slope of the cluster
mass function. We argue that a more economical approach to the problem
would instead involve direct analysis of the upper IMF in clusters
using statistical tests for truncation of the mass function. When such
an approach is applied to data from hydrodynamic simulations we find
evidence for truncated mass functions even in the case of simulations
without feedback. 
\end{abstract}

\section{Introduction}

Does the maximum stellar mass in a cluster depend {\it systematically} on
cluster mass? Here we emphasise the word `systematically'
since we need to distinguish between the pure size of sample effect
(that leads one to expect, on average, a lower maximum mass in a less
populous cluster) and the systematic effect of cluster
mass-dependent truncation of the IMF  explored  by \citet{weidner+kroupa2006}.  A mere size of sample effect would still imply that the IMF
assembled through combining all clusters (the integrated Galactic IMF,
IGIMF) would have the same slope as the universal IMF from which
the stellar content of each cluster was drawn. On the other hand,
cluster mass-dependent truncation of the IMF  within each cluster can result
in an IGIMF with a different slope from the individual cluster IMF:
in particular, if the distribution of cluster masses is strongly biased
towards low mass clusters for which the IMF is truncated at a low value, then
the IGIMF can be markedly steeper than the IMF within each cluster \citep{kroupa+weidner2003}.
This hypothesis would have important implications for how, for example, we
normalise galaxy integrated quantities based on their high mass stellar content
(e.g. supernova rate, blue light) to the total stellar mass/star formation
rate in the galaxy. \citet{weidner+kroupa2005}
have furthermore postulated that if
the cluster mass function is systematically truncated at a mass that depends
on the galaxy mass, then the influence of small clusters is greater
in dwarf galaxies and hence predict a steeper IGIMF in these systems.
It is hard to test the IGIMF hypothesis {\it directly} because, for example,
the observation of a low rate of massive star production (relative
to the available gas mass) may be interpreted either as indicating
a steep IGIMF \citep[so that the the total star formation rate is not
anomalously low:][]{pflamm-altenburg-etal2007}
or else a suppression of the star formation efficiency
(star formation rate per unit gas mass) in that system \citep[e.g.][]{kaufmann-etal2007}. In the
absence of observations that probe the formation of stars on the lower IMF,
both explanations are equally good.

  One can in principle  break this degeneracy by going directly
to observations of young clusters in order to ascertain whether there
is observational support for the effect that underpins the IGIMF hypothesis,
i.e. the systematic dependence of maximum stellar mass on cluster mass
as described above. This hypothesis was originally framed by inspecting
the available cluster data and is thus an empirical hypothesis, without
a quantitative theoretical basis (although one can argue plausibly about
physical effects such as stellar feedback that may give rise to such
behaviour). Therefore in order to establish the need for such a hypothesis
it is first necessary to test the data against the readily quantifiable
`null hypothesis', i.e. that stars are assembled at random from a universal
IMF. Only if the data is significantly discrepant with this hypothesis
should we start to explore and constrain more complicated alternative
hypotheses.

\section{Testing the null hypothesis}

\begin{figure}
\begin{center}
\includegraphics[width=0.5\textwidth]{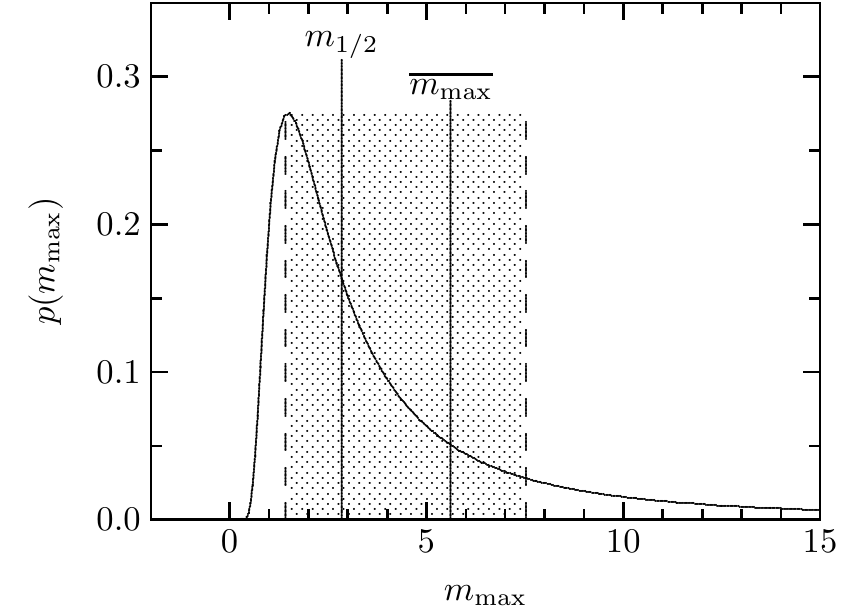}
\end{center}
\caption{\label{mmaxpdfplot}
Probability density of the most massive star, $p(m_\mathrm{max})$ (eq. \ref{mmaxpdf}) for a star cluster containing $n=30$ stars.
Characteristic quantities are the mean, $\overline{m_\mathrm{max}}$, and the median, $m_{1/2}$.
The $1/6$ and $5/6$ quantiles limit the shaded region containing $2/3$rd of the most massive stars.
(Figure taken from \citealp{maschberger+clarke2008}).
}
\end{figure}

  If we pick n stars at random from a universal IMF $f(m)$ (such
that the fraction of stars with masses in the range $m$ to $m + dm$ is
$f(m) dm$) then simple application of binomial statistics
implies that the maximum stellar mass should be distributed according
to the probability density function:

\begin{equation}
  p(m_\mathrm{max}) = n \biggl( \int_{m_{L}}^{m_\mathrm{max}} f(m') dm' \biggr)^{n-1} f(m_\mathrm{max}) \label{mmaxpdf}
\end{equation}

\citep[eq A5 in][]{maschberger+clarke2008} where $m_\mathrm{L}$ is lower limit of the stellar mass function, the (cluster mass independent) minimum stellar mass.
(Note that $m_\mathrm{max}$ is {\it not} the upper limit, $m_\mathrm{U}$ of the IMF).
This can be simply understood as representing  the probability
that one star has mass in the range $m_\mathrm{max}$ to $m_\mathrm{max} + dm_\mathrm{max}$
and the remaining $n-1$ have masses less than this, with there being
$n$ choices as to the identity of the most massive star.  Figure \ref{mmaxpdf}
demonstrates (for the case $n=30$) that this probability density function
is highly asymmetric, with the mean being significantly higher than
the median. This immediately implies that with data drawn
from the null hypothesis the maximum stellar mass is likely (for
the majority of random realisations) to be less than the mean. It
is therefore a mistake to compare the expectation value of the null
hypothesis with sparsely sampled data, since one is then likely
to conclude (erroneously) that the data is systematically low compared
with the model prediction. The correct approach is to compare
observational data with the {\it quantiles} of the distribution.

\begin{figure*}
\begin{center}
\includegraphics[width=0.8\textwidth]{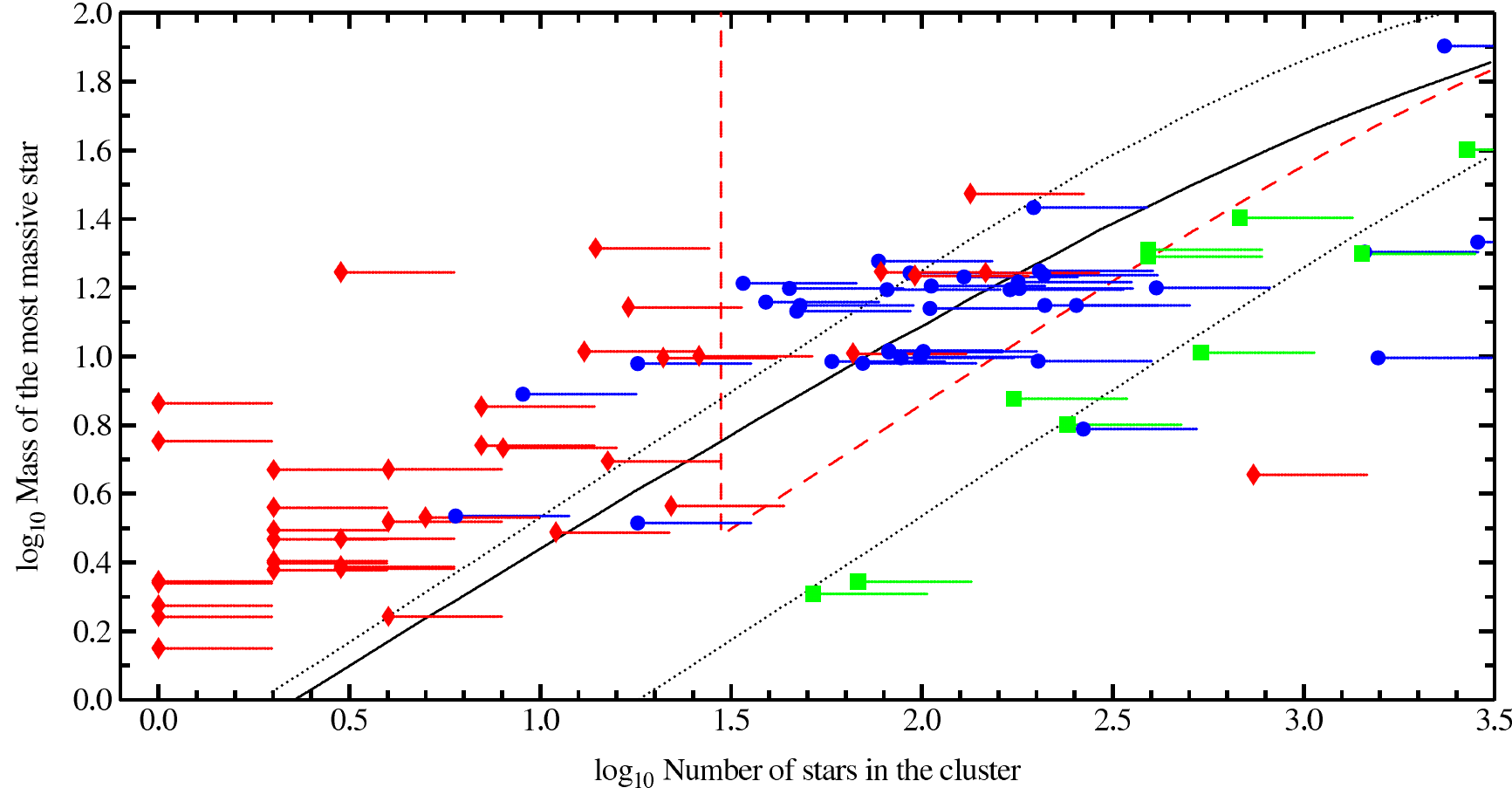}
\end{center}
\caption{\label{mmaxntot}
Mass of the most massive star versus the number of stars in the cluster (for  better visibility, a small random scatter was applied to the (discrete) masses).
The data are collected from the literature, with the main sources Testi et al. ({\color{red} $\blacklozenge$}) and Weidner \& Kroupa ({\color{green} $\blacksquare$}).
The solid line is the mean value of $m_\mathrm{max}$ depending on $n$.
The dotted lines follow the $1/6$ and $5/6$ quantiles, and should confine $2/3$rd of the observed data.
(Figure taken from \citealp{maschberger+clarke2008}).
}
\end{figure*}

  Figure \ref{mmaxntot} represents a recent attempt to collate all available
data on the maximum stellar mass as a function of numbers of stars
in the parent cluster and to compare with the null hypothesis
of random drawing from a parent mass function 
\citep[for details see][]{maschberger+clarke2008}.
 There are several noteworthy
features of the data. Firstly, one is struck by the large range in
maximum mass at a given cluster $n$ (around a factor $5$--$10$) so clearly
one is not looking at a well defined `relation'; on the other hand, the dotted
line shows the $1/6$,$5/6$ quantiles of the null hypothesis which show
that such a range is actually expected in the case of random drawing.
Secondly, the errorbars are large: we have not even attempted to assign
errorbars on the `mass' axis and our  errors in cluster
membership are a notional factor $2$: we have corrected the number of
stars in each cluster down to a common lower mass limit, assuming
a universal (Kroupa) form for the lower IMF but we are obviously unable
to accurately assign the numbers of stars which belong to a given
cluster but whose surface density on the sky is less than the local background
value. This is particularly acute in the case of small $n$ clusters which
are liable to undergo significant dynamical evolution over the (few Myr)
lifetimes of the clusters \citep{bonnell+clarke1999}.
This can result in a significant fraction of original cluster members being
dispersed into the neighbouring field and rendered undetectable, so that
the associated errorbars  are in reality larger for small $n$ systems.
\footnote{It is sometimes stated that one can reduce the problem of
errorbars on cluster membership by plotting this diagram as a function
of cluster {\it mass} rather than $n$: this approach is favoured by
Weidner \& Kroupa, even though it then has the mild disadvantage that
the quantiles have to be computed through Monte-Carlo simulation rather
than using the analytic form, equation (\ref{mmaxpdf}). The argument is that since
lower mass stars contribute less to the total mass than to the total number,
then uncertainties associated with extrapolation to a given lower stellar mass
limit are smaller when expressed in terms of mass rather than number. This
however does not translate into a greater power to distinguish between
rival hypotheses since, just as the total mass is less sensitive
to errors in membership at the lower mass end, so, correspondingly, are
the predicted quantiles of the distribution more sensitive to the
total mass. We have established through Monte-Carlo simulations that, whether
one expresses the independent variable in terms of stellar mass or number,
the  uncertainty associated with where a datapoint is placed with respect
to the quantiles of the null hypothesis is about the same --- in other words,
it does not greatly matter whether one constructs plots as a function of
mass or membership number.}

\begin{figure*}
\begin{center}
\includegraphics[width=0.5\textwidth]{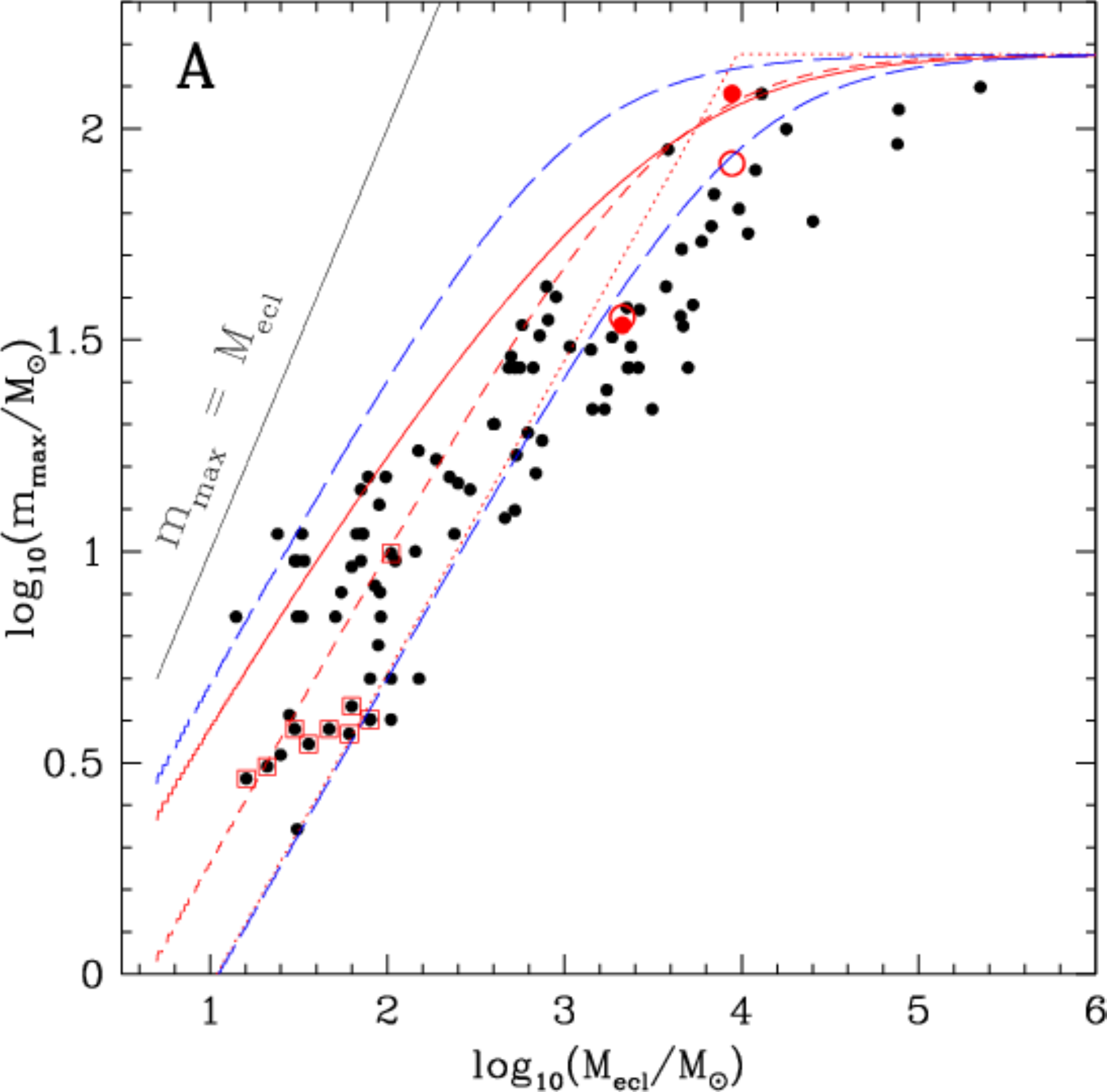}
\end{center}
\caption{\label{mmaxntot_weidner}
Most massive star vs. cluster mass from \citet[][ their fig. 5A]{weidner-etal2010}
The solid line is the mean, short-dashed line is the median and the long dashed lines are the $1/6$ and $5/6$ quantiles for random sampling from the IMF(.
(Figure taken from astroph 0909.1555).
}
\end{figure*}

 Thirdly, one cannot but be struck by the fact that the provenance of the
data determines which region of the diagram it populates. Specifically,
the data of \citet{testi-etal1997,testi-etal1998} lies significantly higher with respect
to the model quantiles than does the other data. This is simply because
the Testi et al. data is assembled by searching for clusters around
known, apparently isolated, massive stars, whereas the other data is
obtained by identifying the most massive member of known clusters.
Clearly, the Testi data will tend to contain objects of relatively
high mass compared with the cluster mass (see also Oey et al, this
volume, for similar results obtained by imaging
OB stars in the Magellanic Cloud). Since our test of the null hypothesis
depends on how the data is distributed in relation to the model quantiles
it follows that the types of objects that we include need
to be representative
of their incidence in the Galaxy. We are of course very
far from this situation when we simply, as in Figure \ref{mmaxntot}, assemble all
measurements from the literature and this means that any conclusions
from plots like Figures \ref{mmaxntot} and \ref{mmaxntot_weidner} must be highly provisional.

   Laying these caveats aside, the  analysis by \citet{maschberger+clarke2008} of the data contained
in Figure \ref{mmaxntot} implies that it is indeed consistent
with the null hypothesis --- we would therefore hesitate to argue
for a cluster mass dependent maximum stellar mass based on this data.

  Figure \ref{mmaxntot_weidner} however tells a different story, being a similar compilation that
extends to richer clusters (i.e. to those containing more than the
few thousand stars contained in Figure \ref{mmaxntot}, from \citealp{weidner-etal2010}). The dashed lines represent
the $1/6$, $5/6$ quantiles of the null hypothesis and show that the
data in the region that overlaps in mass with Figure \ref{mmaxntot} is still broadly
consistent with this (random drawing) null hypothesis. At higher masses, 
however, 
one sees that the data lies progressively lower with respect to the model
quantiles --- i.e. the maximum stellar mass would indeed appear
to be suppressed, in these massive clusters, compared with the
expectations of random drawing.

  In the following section we explore whether we expect the results
contained in Figure \ref{mmaxntot_weidner} to have significant implications for the IGIMF, but
first we need to enquire whether there are any systematics that might
explain trends in the data. We draw attention to the fact that plots like Figures
\ref{mmaxntot} and \ref{mmaxntot_weidner} have to be constructed from clusters that are young enough that
their most massive stars have not yet expired as supernovae. For low mass
clusters, the main sequence lifetime of stars with the rather low
maximum masses expected is fairly long, but as one goes to more massive
systems it follows that one should  include only the  youngest clusters.
Although
Weidner \& Kroupa have been conscientious in applying these age constraints,
one nevertheless has to bear in mind that the ages of young clusters are
not necessarily very accurately known and that age uncertainties could
---in principle--- explain the trend seen in Figure \ref{mmaxntot_weidner}.

\section{IGIMF implications}

\begin{figure*}
\includegraphics[width=0.5\textwidth]{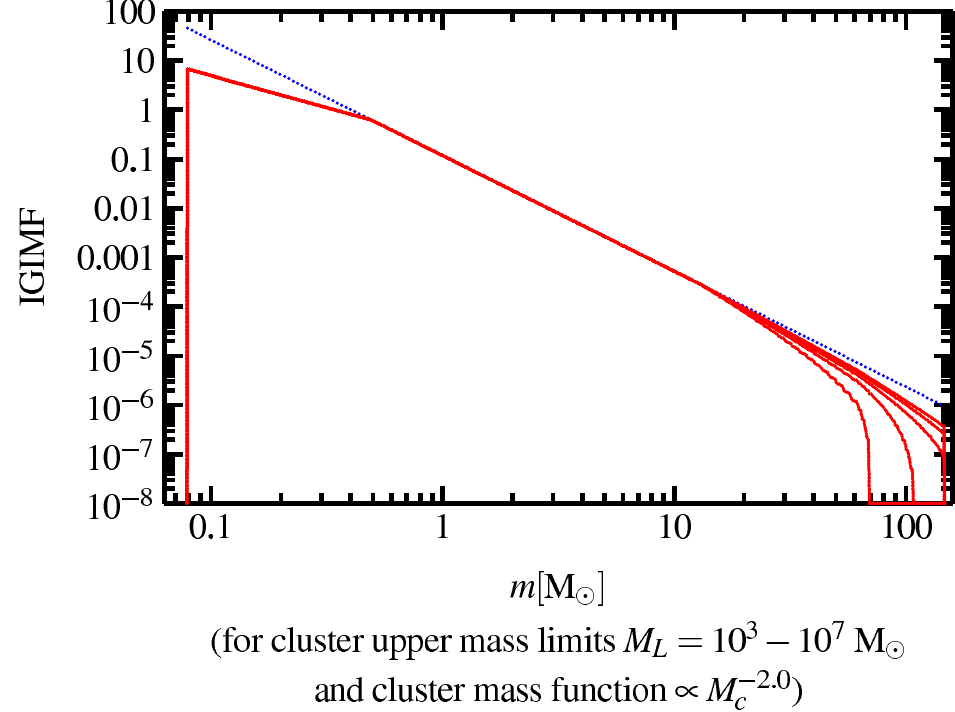}
\includegraphics[width=0.5\textwidth]{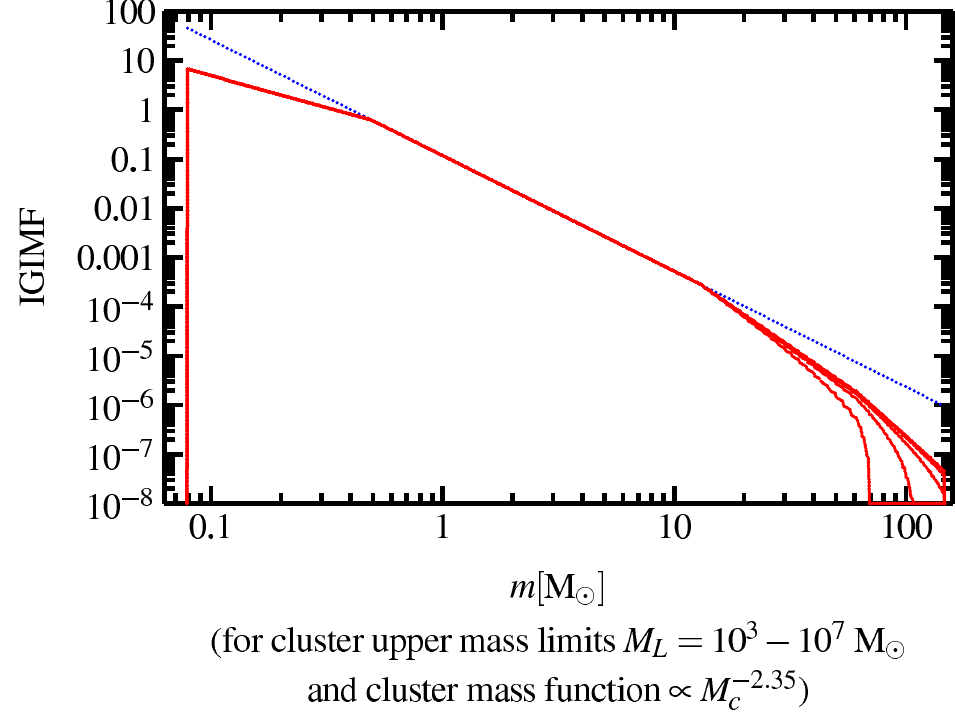}
\caption{\label{igimf}
Integrated galactic IMFs for different values of the upper limit of the cluster mass function (solid lines) in the range $10^3$ to $10^7
M_\odot$. The cluster mass function follows $M_c^{-2}$ (left) and $M_c^-{2.35}$
The dotted line shows a Salpeter stellar mass function $\propto
m^{-2.35}$.
}
\end{figure*}

 We now lay aside all the caveats mentioned above in relation to
Figures \ref{mmaxntot} and \ref{mmaxntot_weidner} and ask whether  ----if Figure \ref{mmaxntot_weidner} were taken  {\it at face
value}--- it would have ramifications for the IGIMF. Since we have no
theoretical guidance as to how the maximum mass should depend on cluster
mass, we just adopt a `toy' relationship which brings the data into
acceptable agreement with the quantiles of the `toy model' (see Figure \ref{igimf}).
Assuming this form, we then consider that the Galactic field is composed
of clusters whose spectrum, by mass, $M_c$, scales as
$\propto M_c^{-\beta}$ and then compute the expected IGIMF.

Unsurprisingly, the result of this exercise depends on $\beta$. If
$\beta$ is $2$ or below, the relative contribution of large and small
clusters as a source of stars of given mass is weighted towards large
clusters, wherein the maximum stellar mass ($m_\mathrm{U}$) is expected to tend to its
global maximum value \citep[here set to $150 M_\odot$: see ][]{weidner+kroupa2004,oey+clarke2005}. Consequently, the IGIMF is indistinguishable from the
IMF in each cluster (here assumed to follow a  Salpeter slope for its
massive stars); this only breaks down if we limit the
maximum {\it cluster} mass to a scale ($< 10^4 \mathrm{M}_\odot$) where the
maximum stellar mass becomes less than $\sim 150 \mathrm{M}_\odot$:  in this
case the IGIMF has a Salpeter slope but is truncated at the maximum
stellar mass that we have assigned  to the maximum cluster mass.

  On the other hand, a rather modest change in $\beta$ changes the
result quite markedly. Figure \ref{igimf} shows the expected IGIMF in the
case that the  slope of the cluster mass function is $\beta = 2.35$: 
this implies a mild bias towards low
mass clusters as the source of Galactic field stars.
In this case one sees the characteristic steepening of the IGIMF
compared with the input (truncated Salpeter) IMF in each cluster.

\section{Discussion}

In some ways Figures \ref{mmaxntot} and \ref{mmaxntot_weidner} represent an exercise that is rather
wasteful of observational data --- i.e. for each cluster we use only
one piece of data (apart from the cluster mass or $n$) namely
the mass of its most massive star. Inevitably this means that
we need a large number of clusters in order to populate this
plane and then we should worry (as discussed above) whether we have
in fact populated this plane in an unbiased way.

 An alternative observational route is however to seek evidence
of truncation of the IMF within individual clusters, through analysis
of the distribution of stellar masses in the high mass tail
of the mass function. \citet{maschberger+kroupa2009}
 present a ready tool for testing for truncation in power law
data which consists of first fitting the data with a truncated power law
and then assessing the significance of the best fit model, i.e. through
testing how discrepant is the data with other hypotheses. This latter
is achieved by means of a stabilised probability-probability (SPP) plot,
which is
closely allied to the Kolmogorov-Smirnov (KS)
test in that it compares the
(stabilised) cumulative distribution of the model with that predicted
by other hypotheses. Here `stabilised' refers to a transformation of the
cumulative distribution in a way designed to achieve uniform variance
at all quantiles in the case of randomly sampled data. This transformation
therefore corrects a well known drawback of the KS test, i.e. that it
is relatively insensitive to differences occurring near the extremes
of the distribution. Obviously, this is a particular drawback if the
feature of interest in the distribution is a truncation at the high mass
end!

The SPP methodology can also be applied to simulation data and has revealed,
in the analysis of stellar
mass distributions generated by the  turbulent fragmentation calculations
of \citet{bonnell-etal2003,bonnell-etal2008}
 that there is some evidence for truncated
mass distributions within individual clusters \citep{maschberger-etal2010}. 
This indeed gives rise to
an (extremely mild) IGIMF effect in the data --- i.e. the IMF of the
stars contained within all the clusters in the simulation volume is
slightly steeper (at a level of about $0.2$ in the mass function
index) than the IMF within individual clusters. What is interesting here
is not the magnitude of the effect (which is far smaller than what
results from the `sorted sampling' algorithms with which \citet{weidner+kroupa2006} illustrate their IGIMF concept) but rather the remarkable
fact that there should be any such effect in simulations which omit
all forms of feedback and in which, therefore, one cannot argue
that any abrupt switch  prevents stars from growing to beyond
a certain (cluster mass dependent) mass. Instead it would seem
that in the calculations, the result is simply due to the fact
that at any point in the simulation
the stars have only had  a finite time in which to acquire mass:
the most massive stars are those that form first and, statistically,
are those that are involved in more successive  cluster
mergers and which  end up in the rich accretion environment of massive
cluster cores. Since stars are often formed in small n groupings, these
groups of massive stars tend to `travel together' up the cluster merger
tree and share similar accretion histories. Thus the most massive
stars are relatively closely  bunched in mass and the resulting IMF
is best fit by a truncated form.

  These calculations are  from representing a `realistic'
picture of mass acquisition on the upper IMF given the omission of
feedback processes; it is interesting that
the history of  cluster and massive star formation in these simulations
is nevertheless such that it gives rise to these
first hints of the reality of an IGIMF effect.

  \section{Conclusions}

  We have shown that current observational data on the relationship
between maximum stellar mass and cluster mass does not strongly
argue for a systematic (non-random) relationship between these
quantities in the case of clusters that number thousands of stars
or less. At higher cluster masses there is an apparent suppression
of maximum stellar mass compared with the expectations of `random
drawing'. We highlight the fact that the conclusions drawn from such
plots are sensitive to the selection criteria involved in acquiring
observational data. We also note that the inference of relatively
low maximum
stellar masses at high cluster mass (Figure \ref{mmaxntot_weidner})
could  in principle be
explained by errors in cluster ages. 

  We briefly examine whether the available data on maximum stellar
mass versus cluster mass could give rise to a significant `IGIMF
effect'. We emphasise that the answer to this question is extremely
sensitive to the power law slope of the cluster mass function, with
the predicted results changing markedly for small variations in this
index around the value of $2$.

  We then argue that it may be more profitable to look for direct
evidence for IMF truncation in massive clusters rather than relying
on one piece of observational information (the mass only of the most
massive star) per cluster. Methods for detecting truncated
mass functions and assessing their significance are available
in the astronomical literature. We show that when such methods
are applied to the output of hydrodynamic simulations of
star cluster formation, we indeed find evidence for truncated
mass functions and a (very mild) IGIMF effect. This truncation
can be explained in terms of the finite time available for
stars to grow by accretion and would probably be more marked 
in simulations that additionally included some form of
dynamical or thermal feedback.


\bibliography{clarke_c}

\end{document}